\documentclass{IETpaper}

\usepackage{graphicx}
\usepackage{pifont}
\usepackage{hyperref}

\newcommand\blfootnote[1]{%
  \begingroup
  \renewcommand\thefootnote{}\footnote{#1}%
  \addtocounter{footnote}{-1}%
  \endgroup
}
\usepackage[hang,flushmargin]{footmisc}

\begin{document}


\title{Eavesdropping Whilst You're Shopping:\\
Balancing Personalisation and Privacy in Connected Retail Spaces}

\author{Vasilios Mavroudis$^{\ast}$ 
\and
Michael Veale$^{\dag}$}

\institution{
$^{\ast}$Dept. of Computer Science, University College London, v.mavroudis@cs.ucl.ac.uk\\
$^{\dag}$Dept. of Science, Technology, Engineering \& Public Policy, University College London, m.veale@ucl.ac.uk}

\maketitle


\begin{abstract}
Physical retailers, who once led the way in tracking with loyalty cards and `reverse appends', now lag behind online competitors. Yet we might be seeing these tables turn, as many increasingly deploy technologies ranging from simple sensors to advanced emotion detection systems, even enabling them to tailor prices and shopping experiences on a per-customer basis. Here, we examine these in-store tracking technologies in the retail context, and evaluate them from both technical and regulatory standpoints. We first introduce the relevant technologies in context, before considering privacy impacts,  the current remedies individuals might seek through technology and the law, and those remedies' limitations. To illustrate challenging tensions in this space we consider the feasibility of technical and legal approaches to both a) the recent `Go' store concept from Amazon which requires fine-grained, multi-modal tracking to function as a shop; and b) current challenges in opting in or out of increasingly pervasive passive Wi-Fi tracking. The `Go' store presents significant challenges with its legality in Europe significantly unclear and unilateral, technical measures to avoid biometric tracking likely ineffective. In the case of MAC addresses, we see a difficult-to-reconcile clash between privacy-as-confidentiality and privacy-as-control, and suggest a technical framework which might help balance the two. Significant challenges exist when seeking to balance personalisation with privacy, and researchers must work together, including across the boundaries of preferred privacy definitions, to come up with solutions that draw on both technology and the legal frameworks to provide effective and proportionate protection. Retailers, simultaneously, must ensure that their tracking is not just legal, but worthy of the trust of concerned data subjects.
\end{abstract}

\section{Introduction}\blfootnote{Proceedings of the PETRAS/IoTUK/IET \emph{Living in the Internet of Things} Conference, London, United Kingdom, 28--29 March 2018.}
The market share of online retailers across many sectors has been steadily growing at the expense of retailers with physical stores. This is partially due to the ease of purchases, but also because of the heavy tailoring and personalisation enabled by digital tracking technologies. However, physical stores are no newbies to customer tracking either. For example, approximately two decades ago, data from Tesco's \textit{Clubcard} scheme in the UK formed the basis for 80,000 variants of a single shot of direct marketing material~\cite{mckinseytracking}. In other instances, retailers have even sought to link credit card details back to the customers' addresses by employing ``reverse append'' practices, bringing a slew of specific legal and legislative responses~\cite{blackmon2014problems}. Today, the proliferation of cheap sensors and ubiquitous smartphones enables the deployment of more advanced behavioural inference systems, some of which aim to drastically alter the physical shopping experience. However, while some personalisation may be desirable---after all, shopping can be a stressful experience~\cite{aylott1998exploratory}---many shoppers may consider it invasive, or feel vulnerable to unethical and illegal decision-making, such as price discrimination based on protected characteristics.

In this paper, we consider the privacy implications of in-store tracking. We first outline applicable technologies, including ambient sensors receiving signals from personal devices (e.g. through WiFi, Bluetooth) or individuals' biometric traces (e.g. appearance, gait), as well as tracking through software installed on individuals' own devices. Following a broad discussion of relevant privacy issues in social and legal context, we then look into mitigation and control practices. In particular, we examine both \textit{purely technical} countermeasures deployable unilaterally by the customer, and \textit{legal obligations} aiming to bind retailers to follow proper data collection and handling practices. Moreover, we highlight times where tracking and its technical and legal countermeasures clash by considering a) the next generation of in-store tracking, typified by the Amazon `Go' proposal, where tracking is required for shopping functionality and b) the technical and legal status of Wi-Fi tracking mitigation. Though these two examples, we identify emerging challenges in both technical and legal governance, and propose promising directions. We conclude by noting that to ensure personalisation remains privacy-preserving and proportionate, researchers must work together to mix different notions of privacy, particularly around biometric tracking, and ensure support for them in legal and technical frameworks, whilst retailers must ensure that any tracking is not just legal, but worthy of the trust of concerned data subjects.

\section{Ends \& Means of in-store Tracking}
The difficulty of tracking in physical spaces caused physical retailers to lag far behind their online counterparts. Yet new tracking products utilising cheap sensors, novel analytics and ubiquitous personal devices have overcome many of these hurdles, meaning potential collection of granular customer behavior data can now be talked about in comparable terms to someone browsing a website. At a basic level, these data can inform operational decisions---stores noticing that people often walk from the ``snacks'' aisle to the ``alcohol'' one, and rearrange the locations of these goods accordingly. While this sounds like useful analytics, in many cases it might be just using an invasive sledgehammer to crack a nut---similar insights may be possible to reach with a range of methods and approaches. More advanced techniques however leave retailers with fine-grained inferences that can be used to take fine-grained action, such as extending personalised offers based on nuanced views of individuals' consumption profiles. Overall, the data collected by in-store tracking systems is applicable to several areas including
\begin{itemize}
\item \textit{Demand Management}: Anticipating general footfall and specific demand for products and services to optimise employee scheduling and stock logistics.
\item \textit{Business Development}: Understanding customer behaviour and habits, so as to make more appealing product offerings or more effective store layouts.
\item \textit{Marketing \& Sales Optimisation}: Using profiling tools for direct marketing or price discrimination.
\end{itemize}

\subsection{Tracking Technologies}

In-store tracking tools gather the quantities and modalities of required data through careful monitoring (i.e. sensing) of customers' behaviour. Sensing in the retail context can be performed in either an active or a passive manner. Active technologies comprise of a network of probes deployed in the retail premises and a sensor carried by the customer. On the other hand, passive technologies rely on sensors that are deployed in the ambient environment and monitor either the customers themselves (e.g. biometric sensors such as facial or gait recognition) or the digital traces left by their personal devices (e.g. Wi-Fi probe-loggers).

\subsubsection{Active Signalling}
This class of customer tracking techniques relies primarily on the users' own devices. Modern handheld devices (e.g. smartphones, watches) come with a variety of sensors and connectivity capabilities useful for such tasks (e.g. Bluetooth or GPS chips). One approach to utilise these sensors is for retailers to develop software (e.g. `apps') and incentivise their installation through discounts or functionality such as information retrieval. Upon installation, the application generally associates itself with one or more environmental triggers. Such a trigger may be the user launching the app, the GPS reporting the coordinates of a brand's branch, or an inaudible signal emitted by the retailer's speakers. Those triggers are then associated with certain actions such as the collection and sharing of real-time data (e.g. location, trigger identifier) with the tracking provider.

\paragraph{Radio Beacons.} These are small devices that emit short-range wireless signals that are then picked up by nearby devices running the retailer's app. In basic deployments, beacons simply echo a unique identifier associated with the in-store area they cover, whilst in more advanced ones they also push real-time information to users' devices. In both cases, the signals emitted by the beacons serve as a trigger for the retailer's app, which then forwards the observed identifiers (along with a timestamp) to the tracking provider's servers. Using this information, the tracking provider can analyse the behavior of each individual (e.g. points of interest, trajectories, product returns), build a profile and even push personalised notifications to their device. The most widespread beacon products use Bluetooth/Bluetooth Low Energy (BLE)---technologies benefiting from both a range ideal for indoor tracking and minimal battery usage on a customer's device. Two notably beacons are Google's Eddystone~\cite{eddystone} and Apple's iBeacon~\cite{iBeacon}, although non-trivial investment is needed to install enough to ensure full space coverage. 

\paragraph{Audiovisual Beacons.} In other cases, retailers employ \textit{ultrasound} based beacons utilising existing speakers to emit high frequency signals inaudible to humans, but which encode a unique identifier~\cite{mavroudis2017privacy}. These can even be embedded in other audible content, like songs. Standard smartphone microphones can capture those high-frequency signals and trigger retailers' apps to report the collected data to the tracking provider. Where multiple devices are being used, such technologies can link the different devices of the user, together in a tracker's database~\cite{zimmeck2017privacy, mavroudis2017privacy}. A similar technique in the visual domain is \textit{light pattern signaling}, where flickering LED lighting unnoticeable by humans can contain unique identifiers able to be captured and decoded by smartphone cameras and software~\cite{saini2012li}. These technologies are not only useful in real-time, but can also apply to data shared by the user later online, such as a video message or post.

\paragraph{Geomagnetic Positioning.} A less widespread technique, geomagnetic positioning uses the Earth's geomagnetic fields and the smartphone's compass to precisely locate individuals in indoor spaces (i.e., 1-2 meters accuracy~\cite{indooratlas2012ambient}). Each building has a unique magnetic ``distortion fingerprint'' occurring from way building materials affect and ``distort'' the otherwise persistent magnetic field generated by the Earth. Those distortion patterns can be mapped to the building's floor plan and track user movements. It should be noted that the compass sensor in Android devices currently requires no access permission, meaning that even where the user has not consented to this tracking on a device level, it could be occurring~\cite{sensors}.

\subsubsection{Passive Signalling}\label{sec:passivesig}
In contrast to active techniques, passive ones do not require any user participation, relying entirely on sensors in the ambient environment designed to pick up biometric signals (e.g. face or gait) or digital traces from devices (e.g. Wi-Fi probe packets).

\paragraph{Wifi Logging.} A popular technique, this involves intercepting, capturing and processing the packets transmitted by Wi-Fi--enabled devices when they are searching for networks to connect to. These packets are broadcast out by smartphones (and all Wi-Fi--enabled devices) to query nearby access points for their name (i.e. SSID) and other characteristics (e.g. encryption ciphers used). They carry the media access control (MAC) address of the device which uniquely identifies it. By monitoring these MAC addresses and the signal strength (i.e. RSSI) retailers can monitor the location of a device and track the customer's behaviour over time. In some cases, triangulation techniques may be used to pinpoint the exact location of the customer. It should be noted that Wi-Fi monitoring does not require any user actions, provided that the customer's device has the Wi-Fi functionality `enabled'.

At first glance, it may appear as if it is hard to link a MAC address to the customer's name. However, retailers employ multiple tracking techniques that may enable them to perform such a pairing. For instance, a MAC address that was tracked to be at a checkout point at a particular time can be easily linked to a customer through the loyalty card used during checkout. For this reason, many mobile operating systems (e.g., Android, iOS) started using ``MAC randomization''~\cite{martin2017study}, where the MAC address reported by the device when probing for networks, is constantly changing in a non-predictable manner. While this technique protects the users from being secretly tracked, it provides no way for them to consent to being tracked by specific vendors---for example, if they are delivering personalisation desired by the user. Later, in Section~\ref{macrandom}, we introduce a new randomisation technique that allows users to opt-in to being tracked by selected vendors whilst they maintain anonymity towards all others.

\paragraph{Imaging Technologies.} These are a class of techniques utilising a network of high-resolution cameras to capture and analyse physical characteristics of shoppers, now more accessible to retailers after a significant drop in equipment prices and a number of breakthroughs in machine vision. A simple use of imaging technologies is for statistical counting, such as measuring footfall in different parts of a store, comparable to entrance turnstiles or laser beams with some additional analytical possibilities. Yet other systems go beyond locating individuals to infer characteristics about them. These range from the relatively straight-forward inference of gender or age to more advanced analysis based on \textit{affective computing}, such as mood, emotion or attention. Further forms of analysis are designed for biometric identification of individuals, from facial recognition to recognition on the basis of other features, such as body shape or gait~\cite{Sightcor, Kairos}. In the case of mood estimation, some emerging technologies claim to track micro-expressions (i.e. very brief, involuntary facial expressions) to better infer the visitor's emotions for the products they encounter while shopping~\cite{Emotient}. Particularly relevant here is eye tracking, a technique to follow a customer's gaze to uncover their aisle browsing habits, what attracts their attention, which visual elements they notice or ignore, and how they interact with products on the shelf (e.g. pick up, return). Until recently eye tracking required that the participants wear special tracking glasses, and hence it was used only in pilot studies. However, Cloverleaf~\cite{Aboutshe} and Affectiva~\cite{shelfpoint} recently introduced shelves that feature micro-cameras capable of tracking eye movement thus enabling retailers to use the technology in real-life deployments.

\begin{figure}
\includegraphics[width=0.5\textwidth]{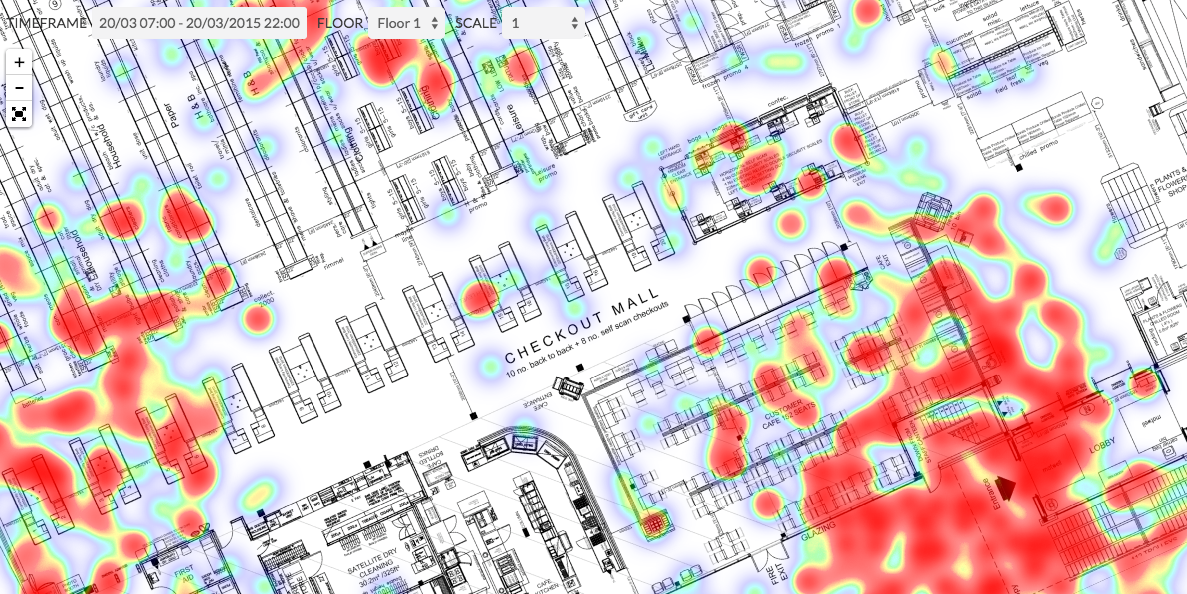}\caption{Heatmap generated by Walkbase tracking product, showing customer movement in a retail store. }
\end{figure}

\section{Privacy \& in-store Tracking}

When considering privacy in retail contexts, we accept that it is an important but contested concept, characterised by ``cacophony, category error, and people talking past each other'' at a range of conceptual and empirical levels~\cite{ohara2016seven}. Consequently, we do not start from one particular notion of it, but from potentially socially contentious data analysis and use in the context of individuals' digital traces in retail outlets and similar physical spaces.

Given this, what private information might be at stake in retail spaces? Assume, without considering the technologies closely at this moment, that your actions and identities as a customer were visible to a retailer. They might include your shopping routine (which may explain something about your job and lifestyle); the volume of your purchases (likely to disclose something about your household); your physical attributes such as appearance or gait (which may betray anything from health data to characteristics such as gender or ethnicity), clothing (which may betray tastes or demographic data); changes in purchase preferences (from which pregnancy or financial turmoil may be inferred); individuals you are commonly seen with, or stop and talk to---and so on. This data impacts upon not just individuals and those close to them, but may also allow further inferences to be made around groups these individuals are part of, such as those sharing a particular health characteristic~\cite{vedder1999kdd}.  While many of these factors may seem at first glance to require detailed study to discern, selected elements of human behaviour are surprisingly predictable, quickly leading to `creepy'--sounding results..

In-store tracking is significantly and increasingly concerning to a large proportion of consumers. According to a 2014 survey of 1,042 US consumers conducted by American consumer feedback company OpinionLab, 80\% of respondents find in-store tracking using mobile phones unacceptable, and 81\% said they don't trust retailers to keep data private and secure~\cite{opinionlab}. A study for the European Commission in 2015 reports that tracking concerns in retail contexts are growing in the UK, with 45\% (of 1,328) of UK residents concerned about tracking via loyalty cards (the third most concerned nation in Europe, up from 36\% in 2005) and 51\% concerned about being recorded in private, commercial spaces (the most concerned nation in Europe, up from 40\% in 2005)~\cite{eurobarometer}. In both these studies, generational differences are not notable: what some have characterised as a lesser emphasis on privacy and data held by younger generations online does not seem to translate into physical contexts.

As will be discussed however, truly avoiding tracking is a difficult task for an individual, if possible at all. In order to manage the issues that arise from the collection and processing of data, we consider two main modes of governance~\cite{diaz2013hero}. The first, privacy-as-confidentiality, seeks to use technological means to ensure minimal information lost or leaked from the individuals carrying out desired tasks. This is characterised by the field of `privacy-enhancing technologies' (PETs), whose researchers utilise primarily cryptographic methods for tasks such as foiling tracking methods or to `have your cake and eat it'---do analysis on whole datasets while learning a minimal amount about individuals within them. The \textit{trust only yourself} mindset of privacy-as-confidentiality is that the world is full of adversaries who are not deserving of trust. PETs researchers assume we need technologies which work even in largely untrusted environments. The second can be described as privacy-as-control: the viewpoint from which data protection law operates. Privacy-as-control seeks to build a \textit{trust the trackers} mindset by seeing them as data controllers with obligations mandated by regulations, including obligations to honour the rights of data subjects whose personal data are being processed by affording them abilities such as to access, erase or port their data, or to object to processing or certain data-driven decision-making. Through this process, it attempts to build social trust in organisations that might otherwise be adversaries, and to shape them into responsible stewards, rather than ruthless exploiters, of data.

Despite the potential for sensitive inferences, we do not deny that some insights could be useful for the purposes of service delivery. Indeed, personalisation is a powerful tool for prioritisation in the information age, and many individuals who dislike shopping with a passion may well find themselves greatly aided by systems that guide their hands and their wallets in physical environments. Personalisation \textit{per se} is not the villain, even though well-known public concerns do exist surrounding extreme personalisation, particularly around the exploitation of data perceived as unjust (e.g. to manipulate customers or to pass to national security) or the inadvertent construction of echo chambers lacking serendipitous exposure to new factors. Indeed, a great deal of work in areas such as web science and human--computer interaction has surrounded how to make better interfaces that adapt more readily to individuals' needs~\cite{brusilovski2007adaptive}. In this paper, we argue that it will be necessary to explore how tracking practices in retail spaces might be governed without relying solely on \textit{either} trust-based control or technologically-assured confidentiality.

\section{Controlling tracking}


\subsection{Trusting only yourself: Unilateral Mitigation}

The first perspective on how tracking can be controlled emerges from the privacy-enhancing technologies (PETs) literature. This attempts to devise technical approaches to mitigate privacy breaches even where the world contains a high proportion of untrusted adversaries. Ideally, such approaches should not restrict a user's ability to make use of a service beneficial to her. Where sensors observe biometric data in public spaces, retaining control is costly. While researchers have developed clothing to `fool' recognition sensors~\cite{sharif2016accessorize}, this places a high burden on an individual simply trying to exercise what in Europe is a fundamental right. Not only are these systems also trying to fool a moving target, the signals given away by individuals attempting to obscure themselves may themselves be used to profile those individuals \textit{en masse}---for example, to deliver higher prices. Where systems track using data from your own sensors, individuals must ensure that no tracking code is being executed on their device. This can be difficult for a number of reasons: individuals can be incentivised to use particular products or services that transmit or enable transmission of data due to market availability, fashion, or perks. Loyalty cards, and more recently, apps, are an example here. The primarily mobile software providers today, Apple and Google, both have interests in physical tracking infrastructure, and individuals have few alternative operating system providers accessible to them on the basis of price, convenience and expertise.

Systems that track using your own sensors, rather than your own signals, such as ultrasound beacons, can be avoided if an individual ensures no tracking code is being executed on your device (or attack being undertaken). This might initially seem like a comparatively simple task — indeed, having tracking code running on your device seems as much a problem of security as of privacy — but it becomes less so when it is considered that these sensors might be enabled by the same apps that users are incentivised to download (and to enable hardware permissions for) due to in-store convenience or savings. When privacy tangibly costs time, money and effort, how possible is meaningful consent?

\subsection{Trusting the trackers: Legal obligations}


It could be argued that a shopping mall is hardly a space where one has a ``reasonable expectation'' of privacy---a legal test present although operationalised differently in both US (Fourth Amendment) and EU (art 8, ECHR) jurisprudence. Indeed, well-known figures wearing oversized sunglasses in public is presumably because these individuals reasonably expect to be looked at and potentially recognised. Yet just as many virtual spaces controlled by private entities have gained a pseudo-public character, nowadays it is more and more possible to capture or infer private information from individuals in public spaces. Much of this personal data (e.g. that broadcast from devices, resulting from conscious interaction with technologies, or biometrically observed) might once have only been stored at home or in a medical record, if at all~\cite{edwards2016privacy}.


While privacy law rooted in fundamental rights (e.g. ECHR art 8 and the US Fourth Amendment) has moved on slightly from considering privacy in public as a total contradiction and imagining that seclusion is suitable choice or alternative, it still remains difficult to convince a court on these grounds~\cite{edwards2016privacy}. Data protection (DP) law on the other hand\footnote{In the European Union, this consists primarily of the GDPR (Regulation (EU) 2016/679 of the European Parliament and of the Council of 27 April 2016 on the protection of natural persons with regard to the processing of personal data and on the free movement of such data, and repealing Directive 95/46/EC (General Data Protection Regulation), OJ 2016 L 119/1.) and the ePrivacy Directive, which is currently being reformed (Directive 2002/58/EC of the European Parliament and of the Council of 12 July 2002 concerning the processing of personal data and the protection of privacy in the electronic communications sector, OJ 2002 L201/37).}, does not distinguish between public and private space, but focuses instead on the difference between personal and non-personal data in determining its applicability. Additionally, it governs public and private actors in (relatively) similar ways, making it a highly unusual regulatory instrument, but consequently a surprisingly wide-ranging one. This places DP law as the core legal structure for concerns relating to retailers' uses of tracking technologies.

The first relevant thing to note is the broadness of the personal data (PD) concept in EU DP law, and the consequences of this for the types of collection activities we have discussed above. While there is no omnibus data protection law in the US, the concepts they do have (`personally identifying information', or PII) tend to rely on an explicit identifier. In the EU, the PD concept is framed around a) any information that relates to b) any natural person identifiable, even indirectly, by some or all of that data. As high dimensional data of many types can be effectively used to `fingerprint' users and single them out in the future, there is no need for data to be stored alongside an explicit identifier such as an individual's name to trigger data protection rights and obligations over it. IP addresses, even dynamic ones, are broadly considered PD in Europe according to the Court of Justice of the European Union (CJEU) (\textit{Breyer}, C-528/14, 2016). The Information Commissioner's Office (the ICO), the UK's Data Protection Authority (DPA), notes that ``using MAC address or other unique identifier to track a device with the purpose to single them out or treat them differently (e.g. by offering specific products, services or content) will involve the processing of personal data''~\cite{ICOwifi}. Some countries have seen relatively high profile investigations into ambient tracking, such as the Dutch DPA's 2015 investigation into the retail Wi-Fi tracking company, \textit{Bluetrace}. The same regulators note that hashing data does \textit{not} render it non-personal, although rigorous hashing (e.g. with a salt that changes daily) is a recommended protection~\cite{ICOwifi}. 

Just because a user is walking around a store does not mean that capturing available data about their characteristics, location or activities is fair game. Where such data relate to mobile devices on the user's person, the relevant law in Europe will depend heavily on the final form of the ePrivacy Regulation which is being redrafted. While the Commission proposal initially permitted the use of Wi-Fi analytics as long as clear notices were put up, the version as amended by the European Parliament (EP) would, unless consent was given, restrict analysis to `mere statistical counting', limited in time and space as strictly necessary for the purpose, which would be deleted or anonymised after the purpose has been fulfilled, with the users given ``effective possibilities to object that do not affect the functionality of the terminal equipment''.\footnote{European Parliament Legislative Resolution on the proposal for a regulation of the European Parliament and of the Council concerning the respect for private life and the protection of personal data in electronic communications and repealing Directive 2002/58/EC, 20 October 2017, art 8.} The crux being that except for simple footfall purposes, Wi-Fi tracking without consent will not be allowed---and such consent must not affect devices functionality, precluding any mechanisms based on turning off device options such as Wi-Fi or Bluetooth.

How might such a consent system work? Assume a retailer seeks to legally use the passive signals sent out by users' devices in a way that would enable the user to be identified---for example to provide some personalised information. Under the legal approach proposed by the EP, the user needs to have opted in in such a way that they can continue to use their device as normal were they to decide against the tracking. Essentially, this requires the data controller to hold a whitelist of device identifiers. As it stands, this is a stronger provision than current regulatory recommendations. The ICO, for example, recommends that physical booths or websites exist where individuals can submit their device to be blacklisted.\footnote{The Future of Privacy Forum, an American thinktank, run a service to syndicate blacklists to providers of Wi-Fi tracking. See~\url{https://optout.smart-places.org/}.} As we note later in section \ref{macrandom}, given the current MAC randomisation of modern smartphones and other devices, maintaining a reliable white-- or blacklist is a much more technically challenging task than it would initially seem. 

Lessons for regulating tensions between consent and tracking can be taken from the governance of the web. Amendments to the ePrivacy Directive in 2011 led to it being known publicly as the `cookie directive', in relation to the way in which is mandated a largely overwhelming and unhelpful confirmation of the placement of cookies across the web. Consequently, to overcome this quagmire of exhausting and ineffective consent-giving, the Article 29 Working Party (also known as the A29WP, a body of EU DPAs regulators with statutory responsibility under the Data Protection Directive to advise on DP matters) has been vocal about both the need to treat web browser `Do Not Track' (DNT) signals as legally binding refusals that override implicit consent, and also for the need for the European Commission to ``promote the development of technical standards for [mobile devices] to automatically signal an objection against [wifi] tracking''~\cite{a29eprivacy}. Yet DNT signals are less straightforward where biometric signals are involved, as individuals usually have little choice about the information they transmitted from their physical person, and how this paradigm might move into the physical world is unclear.


Lastly,\footnote{We consciously omit discussion here about the legality of using fine-grained data for price discrimination. Interested readers are pointed to~\cite{ZuiderveenBorgesius2017}.} under the GDPR it seems likely that organisations engaging in-store tracking of this sort would be obliged to undertake a data protection impact assessment (DPIA). The A29WP, in their draft guidance on this area, suggest that data processing activities involving both ``systematic monitoring of a publicly accessible area on a large scale'' and ``evaluation or scoring [...] especially from [...] locations and movements'' will be considered ``likely to result in a high-risk'' and require a DPIA. If a high risk is determined, then the ``measures envisaged to address the risks, including safeguards, security measures and mechanisms'' must be assessed. In the absence of measures which can mitigate this risk, the data controller must engage in prior consultation with the data protection authority (in the UK, the ICO), who can use a wide range of powers, from investigation to banning processing of this type. While this `soft law' approach in itself is not binding on outcomes, not only does it increase the accountabiilty and points for organisational reflection (for example, reflecting on deployed safeguards), but firms can be fined up to 2\% of global turnover if they fail to undertake a DPIA when it is required.

\subsection{Interim comments}

Trusting yourself and trusting trackers are two approaches to governing privacy. Yet in some cases, they are at tension. Trusting yourself is difficult where institutions have power to incentivise practices that enable tracking (such as through convenience or discounting), where tracking modalities are difficult to switch off or completely hide (e.g. some wireless functions, most biometrics), or where possible modes of obfuscation are costly (e.g. swapping loyalty cards). Yet trusting others is difficult where the relationship is essentially coercive (e.g. where consent or opt-outs are not effective), where keeping track of data controllers is burdensome or overwhelming, or where selective trust is technically challenging (e.g. not using technologies like MAC randomisation opens you up to tracking by non-trusted actors as well)---not to mention the difficulties of effective enforcement of a law which increasingly concerns on \textit{all} actions of all firms whose business models centre on natural persons. 

It is clear that neither technique is a panacea for the social challenges ahead. In the next section, we highlight some of these challenges in context with two cases: the first of a conceptual store run by Amazon which uses tracking for service delivery rather than as a marketing or analytical add-on, the second of the challenges of enabling opt-ins or opt-outs in MAC address tracking, which highlights a tension applicable to many PETs today.


\section{Challenging cases and promising directions}

\subsection{Tracking as a prerequisite to shopping: Amazon `Go'}

Amazon.com, Inc.---the world's largest e-retailer, one of the world's ten largest retailers of any type~\cite{deloitteretail}, and perhaps the global authority on personalised purchasing---has expressed a clear desire to move into physical retail both with its \$14bn purchase of premium supermarket chain Whole Foods Market, Inc. and its hi-tech ``Go'' store concept. The latter generated much publicity as a shop that would sense, using a range of technologies, what you take off the shelf, what you return, and automatically charge you as you leave without tills or cashiers.\footnote{For an Amazon promotional video, see \url{youtu.be/NrmMk1Myrxc} (Internet Archive version: \url{archive.org/details/archiveteam_videobot_twitter_com_805823848050528257}). Note that check-out free stores have been a business trope for some time: see a similar, RFID--powered proposal from IBM in the mid 2000s~\cite{ibmrfid}.} It presented a vision where tracking was not just pervasive throughout a store, but necessary to its operation. While a scaling-up of this concept may yet turn out to be vapourware, a `beta' version in Seattle, WA exists, accessible only to Amazon employees.

Amazon's own promotional material is low on technological details, stating only that the ``checkout-free shopping experience is made possible by [...] computer vision, sensor fusion, and deep learning''. Yet we \textit{can} undertaken some analysis of the types of technologies being considered given highly specific Amazon patents published in recent years relating to checkout-free shopping~\cite{kumar2013detecting}. 

The patent, in addition to the few statements from the firm on the topic~\cite{amazonstatement}, show that Amazon Go is powered not by a tagging technology like RFID but by sensor fusion, profiling and inference. A challenge however, and what appears to already make the proposed technology sufficiently advanced to be `indistinguishable from magic'~\cite{acclarke}, is its ability to detect a user's shopping choices without any explicit affirmative action (such as barcode scanning) beyond what users in-store do already. The patent provides context on how this is done, highlighting that the technologies use a variety of inference-based systems including a) the profiling of the skin colour of an user's hand in order to reidentify them across the store; b) triangulation of audio captured by ambient microphones in order to determine the location of a user; c) triangulation of radio signals from a user's device; d) the use of the GPS chip on a user's device; e) the use of weight sensors and cameras designed to recognise objects on shelves; and lastly f) estimating the most likely option based on ``purchase history and/or what items the user has already picked from other inventory locations'' when analysis from sensors remains uncertain. In addition, facial recognition is suggested as an authentication modality upon entering the store, albeit alongside generic `user provided information'.


In particular, many of the forms of data being processed about individuals (who, by the nature of the system, are identifiable) are what in the GDPR are called \textit{special categories} of data.\footnote{In the GDPR, special category processing is the processing of ``personal data revealing racial or ethnic origin, political opinions, religious or philosophical beliefs, or trade union membership, and the processing of genetic data, biometric data for the purpose of uniquely identifying a natural person, data concerning health or data concerning a natural person's sex life or sexual orientation'' (art 9(1).} Of the above, not only are data concerning race (e.g. the skin tone sensing) singled out, but as is all `biometric data for the purpose of uniquely identifying a natural person'---meaning any of the sensors positioned around the store attempting to identify an individual without using a device on their person may qualify. Where a system decision is made using historical data about picked-up or purchased items, it is very possible that health data is being latently processed---something that was the case in the famous algorithmic `war story' of Target, whose loyalty card information was reportedly used to detect pregnancy among its customers~\cite{nyttarget}. Furthermore, not only can recorded audio data easily contain special category data such as political opinions, it is also rarely considered proportionate in a public space by regulators~\cite{ICOcctv}. 

In the GDPR, all personal data processing requires a valid legal ground. In the Amazon Go case, consent---only one legal ground of many\footnote{Many, particularly in the media, misleadingly believe and incorrectly report that the GDPR obliges data controllers obtain consent for data processing---this is not true.}---would not be easy to rely on. Firstly, any consent must be \emph{freely given, specific, informed and unambiguous}. Given that any user in this store seemingly has no choice (see below) \textit{but} to be tracked, it is unlikely that this would be freely given. The GDPR explicitly notes that consent is unlikely to be considered as freely given when the ``performance of a contract, including provision of a service is conditional on consent to the processing of personal data that is not necessary for the performance of that contract'' (art 7(4)). Ubiquitous in-store surveillance---which in theory might quickly scale to many stores, rather than specific ones we might think of as novelties---does not seem ``necessary'' to the purchasing of groceries, which would be the nature of the contract or service in this situation.\footnote{While we do not expand upon it in this paper, we note that this hits against the hotly contested legal area of ``mall access litigation'' which is increasingly a symbolic test of the limits of private ownership and the potential for exclusion in an increasingly ``pseudo-public'' patchwork of urban spaces~\cite{gray1999civil,mac2012virtual}.} Because of this, data controllers would generally rely on either ``legitimate interest'' or ``for the performance of a contract'' grounds to justify processing. Yet in this case, if the controller accepts special categories of data are being processed, neither of these are options under the law.

As explicit consent is the legitimisation ground likely to be required in this case, it is key that when consenting, users can effectively refuse consent to the processing of at least the special categories of data. Only then can the consent be considered freely given, and those that wish to consent to such processing can be considered to have done so in the eyes of the law. But to have controller \textit{avoid} capturing and processing special categories of your data in a building filled with ambient sensors is, of course, easier said than done. As discussed above, unilateral methods to avoid tracking are, in general, practically unworkable when biometric data is being captured in physical environments. Some trust in the data controller to implement privacy-preserving techniques during processing is likely necessary. Yet with conventional methods, even to provide your profile such that you can be robustly `ignored' by the system would require the processing of special categories of data. The UK's Data Protection Bill, under debate in Parliament at the time of writing, has no derogations relating to the use of special category data in order to opt-out from further collection,\footnote{Such a derogation would be possible within the context of art 9(g), GDPR.} and indeed it is easy to see the wholesale collection of such data being of concern to privacy activists even if not used for other purposes, due to the need to keep it stored securely. In this case, what is required is a mix between confidentiality and control: the data controller must learn enough about you so that they can exclude you from the system, but not enough so that either the user is uncomfortable, or that they can be considered to be processing special categories of data. For example, biometric data might be subject to a hashing mechanism or similar set-up to allow sensors to locally verify whether an individual being sensed was on an opt-out list and avoid the recording of data about them may serve such a purpose. Such technologies have been proposed in the context of face recognition~\cite{gomez2014protected}, yet the nature of sensor fusion makes this task challenging. Indeed, there are legislative moves towards making `Do Not Track' signals legally binding,\footnote{See the discussion in~\cite{a29eprivacy} in relation to the draft ePrivacy Regulation.} so working out how to send these effectively and privately in physical, multi-modal environments should be a technological priority for research in the years to come.

\subsection{Privacy by Default with Selective Consent} \label{macrandom}
\begin{figure}\label{fig:triangulation} \centering
\includegraphics[width=0.28\textwidth]{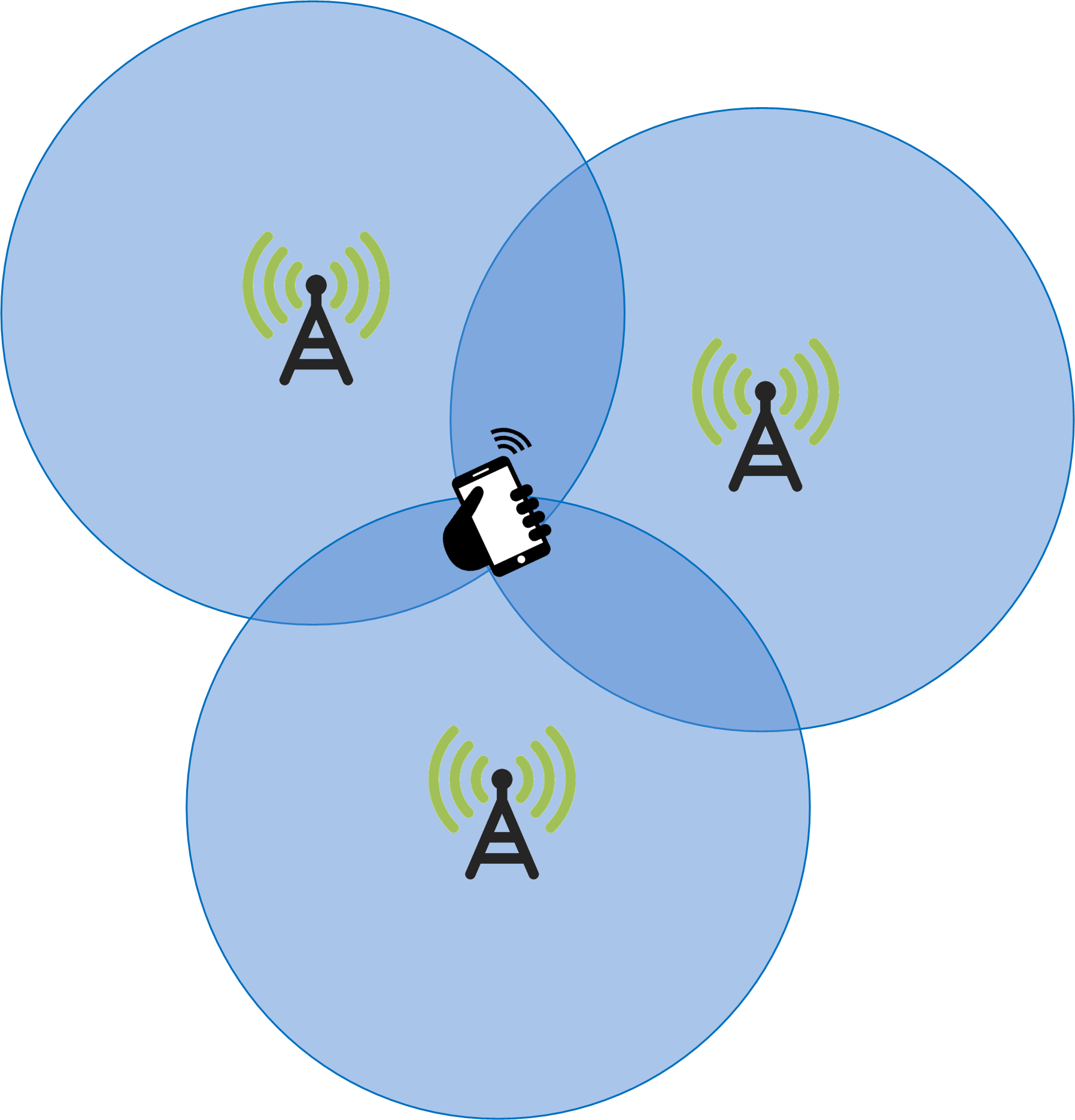}\caption{Triangulation techniques are used by retailers to infer the exact location of a mobile Wi-Fi--enabled device.}
\end{figure}

As outlined in Section~\ref{sec:passivesig}, Wi-Fi probe packets are used to monitor the movement of customers in indoor spaces. From a technical perspective, this is achieved by capturing the \textit{probe request frames} sent by customers' devices. These packets are broadcast by smartphones, laptops (including many on standby mode), and other devices that are not currently connected to a Wi-Fi network. Every few seconds, the device broadcasts multiple such packets, querying networks it has connected to before. The packets have multiple fields which, among others, specify the queried network's name (i.e. SSID), the supported communication rates and the device's factory-assigned, unique identifier (i.e. MAC address).  The probe loggers deployed around the store capture these packets and match those originating from the same sender (i.e. same MAC address). While no location information is embedded in the probing frames, Wi-Fi transmissions have a well-defined range and the exact location of the device can be inferred using triangulation (Figure~\ref{fig:triangulation}).

To prevent this type of tracking, many major manufacturers and operating systems (e.g. Android, iOS) have implemented protocols to generate temporary, randomized MAC addresses that differ from the factory-assigned one~\cite{martin2017study}. This mitigates tracking, as the probe packets report the periodically-changed, pseudonym addresses instead of the real one---although researchers note that such mitigation is often only partial, and significant re-identification can remain possible~\cite{vanhoef2016mac}. However, this partial technical solution had an unexpected side-effect. It prevented tracking providers from allowing customers to either opt-out or opt-in to tracking, and as a result they lost the ability to either manage their own risk, or consent to service provision which might have Wi-Fi triangulation as a useful data collection modality. As the pseudonymous MAC addresses are picked at random and are changed periodically, the tracker does not know if an observed MAC address belongs to a user that asked not to be tracked or not, meaning the the blacklists regulators recommend~\cite{ICOwifi} are ineffective almost instantly after a user has committed to them.

We outline here an approach to be developed in further work to illustrate the potential for trade-offs between personalisation and privacy-by-default. We propose a hybrid randomisation scheme which enables customers to selectively allow vendors to track them, whilst they maintain the ability to unilaterally opt-out at any time. From a technical perspective, we exploit the fact that Wi-Fi--enabled devices broadcast probe frames for each of the networks they have been connected to, in the recent past. When a user opts-in to being tracked by a specific vendor, then the vendor provides them with a ``Vendor ID'' and a unique secret ``Seed''. The user then sets up a new network connection with the Vendor ID as the SSID. From now on, the device will also broadcast probes for the vendor's network, when seeking for networks to connect to. The seed is used to enable tracking from that particular vendor. More specifically, instead of generating new MAC addresses at random, a pseudorandom number generator is used to generate new MACs based on the secret seed, and the time and date (i.e., UNIX timestamp). This enables a vendor who knows the seed for a particular customer to link their pseudorandom MAC address back to them, while the MAC will appear random to everyone else. As an alternative, the pseudorandom identifier can be stored in any other field of the probe packet. To opt-out, the user simply removes the network details from the device's wireless networks list and the device will stop broadcasting for that network. Such permissioning could be managed through user apps or through the native provisions of operating systems designed to enable data protection by design.

Overall, our randomisation scheme enables a more flexible approach to customer tracking as it enables the user to decide for themselves, and gives them complete control over their data and who they are sharing them with. It is a simple specification, but one whose logic could be applied more broadly across different types of privacy enhancing technologies to enable more flexibility in the range of trade-offs deployed.

\section{Concluding remarks}
While the idea of personalisation in retail is not new, the volume of heterogeneous information sources needed to build accurate customer profiles and make precise inferences hindered its adoption until recently. This drastically changed with the proliferation of IoT and other connected technologies that can be used as tracking sensors for individuals. These advancements enabled retailers to track and profile their customers on an individual level, thus mimicking practices used by online stores. Proportionate tracking for better service provision to customers may not be a bad thing in and of itself, if customers have reliable controls over the extent to which knowledge about them is accessible and utilised, and technologies to help this minimisation and purpose limitation are put in place. Yet without safeguards, ubiquitous tracking in physical spaces pose various severe challenges social challenges---particularly, as we highlight, to privacy.

In this paper, we have argued that these challenges cannot be realistically solved by either law or technology alone, which tackle different parts of the problem, have their own relative strengths, and without co-ordination can sit at tension. Technical approaches can assure desired outcomes in a limited set of situations, and if implemented correctly can be powerful privacy tools. Yet there is only a limited amount that users can unilaterally achieve, and many privacy preserving approaches require the buy-in of data controllers too. These cooperative efforts are not trusted by laypeople because they have e.g checked the cryptography for mathematical soundness themselves, but because they trust social institutions which they deem credible and legitimate. Legal approaches allow for redress and enforcement, and are potentially more accessible in relation to physical trackers with a local legal presence compared to faceless online data brokers, yet not only is the law difficult to access for users, but regulators have traditionally been underfunded and outgunned in relation to the scale of digital societal transformations. Both approaches are hard to implement where individuals have incentives to hand away data for economic purposes or for convenience. As the European Data Protection Supervisor argued

\begin{quote}
There might well be a market for personal data, just like there is, tragically, a market for live human organs, but that does not mean that we can or should give that market the blessing of legislation. One cannot monetise and subject a fundamental right to a simple commercial transaction, even if it is the individual concerned by the data who is a party to the transaction.~\cite{edps}
\end{quote}

We believe, in line with the trajectory of European data protection law, that individuals should not be pressured to hand over personal data except for services for which those data are strictly necessary to provide (e.g. personalisation authorised by a user). Consequently, we propose a combination of technical countermeasures and the utilisation of GDPR rights and obligations that enables individuals to better exert control over these technologies, including, if they want, to refuse to be tracked entirely.

Biometric tracking presents a considerable challenge for this area, and further research will be needed to work out how consent can function with ambient sensors collecting sensitive data---if it can function effectively at all. Indeed, as we note with the Amazon Go concept store, some business models are now predicated on collecting sensitive categories of data, such as ethnicity, and using them as a necessary part of service delivery. This is difficult to reconcile with the law as is stands, as we caution that such approaches which by their nature do not allow opting out of tracking without opting out of service use, could pressure individuals into sacrificing fundamental rights for economic reasons. Regulators and technologists both need to consider how potentially `seamless' technologies can remain proportionate going forwards.

Privacy and data protection by design of all types is hugely important to ensure that infrastructures cannot be misused. Law is only law if it can, in theory, be broken. Relying on trust alone is problematic by nature in respect of this, as society often only grants license to install such systems on the basis that their purpose is limited. But ignoring trust is equally problematic, and pretending what individuals and institutions believe is not important risks wide mistrust of a huge array of technologies across the board regardless of application---trust which can be difficult to recover. The act of personalising is not inherently privacy-invading when done correctly, but if it is undertaken poorly and recklessly, personalisation risks becoming synonymous with exploitation. Balancing personalisation and privacy requires giving individuals the practical ability to determine how to be seen by trackers, the assurance that any data provided for a personalisation service is proportionate and necessary, as well as both control throughout the process and technical assurances that such control is as sound as possible give the state-of-the-art. This is daunting, but certainly within the reach of resourceful researchers and practitioners, and we should strive to achieve it.

\section*{Acknowledgements}
Vasilios Mavroudis was supported by the European Commission (H2020-DS-2014-653497 PANORAMIX project). Michael Veale was supported by the EPSRC (grant no EP/M507970/1).

\bibliographystyle{plain}
\bibliography{bibliography.bib}

\end{document}